\documentstyle[12pt]{article}
\begin{document}
\begin{titlepage}
\pagestyle{empty}
\baselineskip=21pt
\rightline{SUSX-TH-95/18}
\rightline{May 1995}
\vskip .2in
\begin{center}
{\large{\bf Relaxation mechanisms: From Damour-Polyakov to Peccei-Quinn
\\}}
\end{center}
\vskip .1in
\begin{center}
C.E.Vayonakis \footnote {On leave of absence from Physics Department,
University of Ioannina, GR-451 10 Ioannina, Greece}

{\it Physics and Astronomy Subject Group,}

{\it University of Sussex, Brighton BN1 9QH, U.K.}

\vskip .1in

\end{center}
\vskip .5in
\centerline{ {\bf Abstract} }
\baselineskip=18pt
The relaxation mechanism of Damour-Polyakov for fixing the vacuum
expectation value of certain scalar fields (moduli) in string theory
could provide a convenient framework for the Peccei-Quinn relaxation
mechanism and remove the narrow ``axion window''.

\noindent

\end{titlepage}
\newpage
\baselineskip=18pt

\def\la{{{\lower 5pt\hbox{$<$}} \atop {\raise 5pt\nbox{$\sim$}}}~}
\def\ga{{{\lower 2pt\hbox{$>$}} \atop {\raise 1pt\hbox{$\sim$}}}~}
\def\mtw#1{m_{\tilde #1}}
\def\tw#1{${\tilde #1}$}
\def\beq{\begin{equation}}
\def\eeq{\end{equation}}
\def\lessim{\lower0.6ex\hbox{$\,$\vbox{\offinterlineskip
\hbox{$<$}\vskip1pt\hbox{$\sim$}}$\,$}}
\def\grtsim{\lower0.6ex\hbox{$\,$\vbox{\offinterlineskip
\hbox{$>$}\vskip1pt\hbox{$\sim$}}$\,$}}

\noindent{\bf I.\quad The Peccei-Quinn mechanism}
\medskip
\nobreak

Relaxation mechanisms, by which some physical parameters can be dynamically
relaxed to their (presumably small) values, are not unknown in
physics.

In particle physics the most famous example is the Peccei-Quinn
mechanism for solving the strong CP problem. Quantum chromodynamics
(QCD), the $SU(3)_C$ gauge theory of strong interactions, allows a
topological term
$$
L_{\theta} = {\theta \over {32{\pi}^2}} {G_{\mu\nu}}^{\alpha}
{\tilde{G}}^{\alpha \mu \nu}
\eqno{(1)}$$
If $\theta \neq 0$, this term induces through non-perturbative
QCD-instanton effects violations of P and CP in the strong interactions.
However, no such violations have been observed and the upper limit on the
electric-dipole moment for the neutron requires $\theta \lessim
10^{-9}$. The strong CP problem is the question why the parameter
$\theta$ is so small. The Peccei-Quinn mechanism \cite{pq} is based on
the idea of making the parameter $\theta$ a dynamical field $\theta(x) =
{\alpha(x) \over f_\alpha}$, where $\alpha (x)$ is a dynamical
pseudo-scalar field called axion and $f_\alpha$ ( known as decay
constant) is the vacuum expectation value at which the global
Peccei-Quinn $U(1)_{PQ}$ symmetry is spontaneously broken. The axion
field is taken to reside in the phase of a standard-model (SM) ($SU(3)_C
\times SU(2)_L \times U(1)_Y$)-singlet complex scalar field $\varphi =
\frac{f_\alpha}{\sqrt2} e^{i\alpha/{f_\alpha}}$ with potential
$$
V(\varphi) = {\lambda}{\left(|\varphi|^2 - {{f_\alpha}^2 \over 2}
\right)^2}
\eqno{(2)}$$
The axion $\alpha$ corresponds to the flat $\theta = {\alpha \over
f_\alpha}$ degree of freedom and would have been massless (true
Nambu-Goldstone boson) if there were not non-perturbative effects that
make QCD depend on $\theta$ and break explicitly the global $U(1)_{PQ}$
symmetry at the scale $\Lambda_{QCD}$. These effects produce an effective
potential
$$
U(\theta) =  U({\alpha \over f_\alpha}) = {\Lambda_{QCD}}^4 (1 -
\cos{N_{dw}\theta}) = {\Lambda_{QCD}}^4 \left(1 - \cos{N_{dw}{\alpha
\over{f_\alpha}}}\right)
\eqno{(3)}$$
where $N_{dw}$ is an integer depending on the theory and associated with
domain walls \cite{sikivie}. One usually takes $N_{dw} = 1$
\cite{vs}. The potential (3) allows $\theta$ to relax to zero
dynamically thus solving the strong CP problem. Moreover, the axion
acquires a mass ( it becomes a pseudo-Nambu-Goldstone boson ), which
scales like ${f_\alpha}^{-1}$ : $m_\alpha \sim
\frac{\Lambda_{QCD}^2}{f_\alpha} \sim 10^{-5}\rm eV
\frac{10^{12}\rm GeV}{f_\alpha}$. Its couplings also scale like
${f_\alpha}^{-1}$. Thus, a very light axion (very large $f_\alpha$) is also
very weakly coupled, hence the term invisible \cite{invisible}.

Various arguments constrain the axion mass $m_\alpha$ and the breaking
scale $f_\alpha$ to lie in a very narrow window. In fact, searches for
the axion in high-energy and nuclear physics experiments \cite{kim} and
astrophysical considerations \cite{turner} require $m_\alpha \lessim
10^{-3}\rm eV$ ( $f_\alpha \grtsim 10^{10}\rm GeV$ ). On the other hand,
by asking that axions ( through their coherent oscillations around the
equilibrium value $\theta = 0$ ) do not overclose the universe, the
famous cosmological constaint $m_\alpha \grtsim 10^{-5}\rm eV$ ( $f_\alpha
\lessim 10^{12}\rm GeV$ ) is obtained \cite{pww}. Moreover, since the
Peccei-Quinn symmetry breaking involves the spontaneous breaking of a
global U(1) symmetry, strings are produced \cite{ve,vs}, which decay by
radiating (among other things) axions. It was argued \cite{davis,vs}
that this could strengthen the cosmological constraint $m_\alpha \grtsim
10^{-4}\rm eV$ ( $f_\alpha \lessim 10^{11}\rm GeV$ ), although this is
a matter of debate \cite{hs}. There remains, thus, a narrow ``axion
window'' $10^{-5}\rm eV \lessim m_\alpha \lessim
10^{-3}\rm eV$ ( $10^{10}\rm GeV \lessim f_\alpha \lessim 10^{12}\rm GeV$
), to which existing projects of experimental search for axions are oriented
\cite{rs}.
\bigskip

\noindent{\bf II.\quad The Damour-Polyakov mechanism}
\medskip
\nobreak

In superstring theory the Damour-Polyakov mechanism \cite{dp} offers
another example of a relaxation mechanism by which various moduli fields
$\Phi$ are attracted towards their present vacuum expectation values due
to string-loop effects. The idea is that non-perturbative effects, associated
with higher genus corrections, may naturally generate different
non-monotonic coupling functions $B_i(\Phi)$ of $\Phi$ to the other
fields, labelled i, of the form
$$
B_i(\Phi) = e^{-2\Phi} + {c_0}^{(i)} + {c_1}^{(i)} e^{2\Phi} + ...
\eqno{(4)}$$
Note that in the case of the dilaton such a coupling function already starts at
the tree level (the first term in equation (4)), whereas for the other
moduli fields it will arise at the one loop level and beyond.
Under the assumption that the different coupling functions $B_i(\Phi)$
have extrema at some common point $\Phi = \Phi_m$ (which is guaranteed
if they coincide $B(\Phi) \equiv B_i(\Phi)$), the expanding universe
drives the vacuum expectation value of $\Phi$ towards the value
$\Phi_m$ at which its interactions with matter become very weak \cite{dp}.

In fact, under an appropriate rescaling $\Phi \rightarrow \phi =
\phi(\Phi)$, all relevant couplings are $\propto \delta \phi$, where
$\delta \phi = \phi - \phi_m$ is the relaxation shift of the moduli
field towards $\phi_m$. Deviations from general relativity are
proportional to $(\delta \phi)^2$ and the present high-precision tests
of the equivalence principle require $\delta \phi \lessim 10^{-6}$. All
these deviations have been actually estimated in this scheme to be
sufficiently small at the present cosmological epoch \cite{dp}.
Additional astrophysical and cosmological considerations may require a
further strong suppression \cite{cev}.
\bigskip

\noindent{\bf III.\quad Implications of an inflationary era}
\medskip
\nobreak

Inflation \cite{guth} has been extensively discussed in the past in
relation with the Peccei-Quinn mechanism. One possibility is to have
(either no inflation at all or) the Peccei-Quinn symmetry breaking down
after inflation. The narrow ``axion window'' mentioned above is now relevant
and it remains to be seen if it is realized in nature. The most serious problem
in this case is the axionic domain
wall problem \cite{sikivie}. For $N_{dw} = 1$ the
problem does not exist, since then the domain walls are bounded by
axionic strings and this can lead to their decay before they dominate
the universe causing, thus, no considerable cosmological effects \cite{ve}.

On the other hand, if the Peccei-Quinn symmetry breaks down before the
end of inflation and the reheating temperature after inflation is lower
than the Peccei-Quinn symmetry breaking scale, then the domain wall
problem disappears, since the domain walls problem are inflated away.(It has
been argued \cite{ll} that quantum fluctuations of
the axion field during inflation may still lead to a domain wall problem
even for $N_{dw} = 1$, but again domain walls are inflated away except
if they are produced late enough.)\footnote {Although there could exist
solutions of the axionic domain wall problem relying purely on particle
physics, see e.g. ref. \cite{ls}, or cases with the cosmological
constraints on the axion mass relaxed simply due to possible physical
processes of a large entropy increase at late stages in the evolution of
the universe, see e.g. ref. \cite {lps}, inflation being an influential
idea per se remains the most appealing solution in many particle physics
problems related in one way or another to cosmology.}

If the Peccei-Quinn symmetry breaking occurs during inflation, anthropic
principle arguments have been invoked to relax the cosmological
constraint on the axion mass and open the ``axion window'' \cite{pi}.
However, it was subsequently argued \cite{tw} that, even with inflation,
it is rather difficult to avoid the constraint $m_\alpha \grtsim
10^{-5}\rm eV$ ( $f_\alpha \lessim 10^{12}\rm GeV$ ), mainly due to
considerations of isocurvature density perturbations produced during
inflation by quantum fluctuations of the axion field \cite{twz}.
(Nevertheless, it was pointed out \cite{linde} that inflationary
models exist where the constraint $m_\alpha \grtsim 10^{-5}\rm eV$ (
$f_\alpha \lessim 10^{12}\rm GeV$ ) may still be avoided.)

Coming now to the Damour-Polyakov mechanism, inflation may be a necessity
\cite{cev}. A detailed analysis of a primordial inflationary era within
this mechanism has been done in ref. \cite{dv}. It was there shown that such
an era could easily solve the Polonyi-moduli problem \cite{cfkrr} and,
moreover,
the produced quantum fluctuations $\delta \phi$ of the relevant moduli fields
during this era are naturally compatible with the observational requirements
from general relativity.
\bigskip

\noindent{\bf IV. A possible scenario}
\medskip
\nobreak

We will now present a possible scenario in which the Peccei-Quinn mechanism is
realized
in a superstring-theory context with the
Damour-Polyakov ans$\ddot{a}$tz and examine the consequences.

First, we notice that in superstring theory with N = 1 supergravity the
potential axions are massless scalars closely connected with the anomaly
cancellation mechanism \cite{gsw}. They originate from the two form B
residing in the supergravity multiplet. We encounter a model-independent
scalar zero-mode (it arises in a way that does not depend on the details of
compactification), as well as a model-dependent one. They exhibit
couplings to tr$G\tilde{G}$ and give, thus, a four-dimensional scalar
behaving as an axion $\alpha$. The tr$G\tilde{G}$ coupling is the dominant term
violating the axionic Peccei-Quinn symmetry non-linearly realized :
$\alpha \rightarrow \alpha$ + c, c = constant. (Cosmological
implications of domain walls in superstring theory have been discussed
in ref. \cite{cd}.)

The important thing for us is that a potential axion field resides among
the moduli fields of a superstring theory. So, for that the
Damour-Polyakov ans$\ddot{a}$tz is applicable. Then, we can imagine the
following picture.

The Peccei-Quinn symmetry is broken when a SM-singlet
complex scalar field acquires a vacuum expectation value $\sim f_\alpha$
minimizing a potential as in (2). In a superstring theory $f_\alpha$ is
naturally of the order of the Planck mass $\sim M_P$. The $\theta =
{\alpha \over f_\alpha}$ degree of freedom is a flat direction (flat
directions naturally arise in the effective supergravity theories
anyway) and corresponds to the axion field $\alpha$. Being a moduli
field, this degree of freedom develops a coupling function $B(\theta)$
$\grave{a}$ la Damour-Polyakov as in (4) (we consider a common coupling
function). The Peccei-Quinn symmetry is broken before the end of
inflation, which is driven by some scalar field $\sigma$ interacting
with the Peccei-Quinn field (it could be that the Peccei-Quinn field
itself is the inflaton, as in `` modular cosmology'' \cite{bbsms}; we
assume here that the inflaton is some other field). Note that, although
the non-vanishing vacuum energy present during inflation can lift the
flat directions of the effective supergravity theory, this is not
necessarily the case \cite{gno}. Then, as explained in ref.
\cite{dv,cev}, at the end of inflation the dynamical variable $\theta$,
irrespective of its initial value, is quicly relaxed extremely close to its
equilibrium point
$\theta_m$ :
$$
\delta \theta = (\theta - \theta_m) \sim e^{-c H\tau} \lessim 10^{-30}
\eqno{(5)}$$
for $c \sim O(1)$ and $H\tau \grtsim 70$, where H is the approximately
constant Hubble parameter during the slow-roll period $\tau$ of
inflation. The equilibrium value $\theta_m$ is naturally guaranteed to
be $\theta_m = 0$ if there exist a discrete duality symmetry. Discrete
duality symmetries are known to hold for moduli fields and, in fact,
motivate the Damour-Polyakov mechanism.

The result is that in this case the axion angle $\theta$ in the early universe
- the
so-called ``misalignment'' angle - is quickly settled down to $\theta =
0$ at the end of an inflationary era within a causal region from which
our entire presently observable universe has originated.

However, in addition there are quantum fluctuations arisen at the late
stages of inflation. They set an absolute minimum to the effective
misalignment angle and give rise to isocurvature axion fluctuations
(fluctuations in the local axion-to-photon ratio) \cite{twz}, which
later evolve into density perturbations of the same maginitude leading
to fluctuations in the temperature of the cosmic microwave background
radiation (CMBR). The relevant quantum fluctuations in the axion field in the
scheme under discussion can be extracted from ref. \cite{dv}. The
largest possible ones have a size
$$
\delta \theta \sim 10^{-7} 10^{-13\kappa} \left(\frac{10^5
H_*}{M_P}\right)^{1 - \kappa/4}
\eqno{(6)}$$
where $H_*$ is the expansion rate at the end of inflation $t_*$ and
$\kappa \equiv -B''(\theta_m) / B (\theta_m)$ is a parameter
of the model expected to be $\lessim 1$. For $H_* \lessim 10^{-5}M_P$
(larger values of $H_*$ lead to excessive amount of relic gravitational
waves), the fluctuations (6) induce anisotropies of the CMBR temperatures
(order of magnitude estimates) ${\delta \theta} / {\theta} \sim {\delta \rho} /
{\rho} \sim {\delta T} / T$ safely
smaller than the experimental constraint ${\delta T} / T \lessim
10^{-5}$.

After inflation, the universe is left with a misalignment angle very
close to zero. The thermalization temperature is estimated \cite{kls} to be
$T_*
\lessim {N_*}^{-1/2} \left( H_* M_P \right)^{1/2}$, where $N_*$ is the
number of the effective relativistic degrees of freedom. A
representative value is $H_* \sim 10^{-7} M_P$, for which $T_*
\lessim 10^{15}\rm GeV$. The QCD-instanton effects are not operative until
sufficiently small
temperatures $T \sim \Lambda_{QCD}$, at which the field $\theta$ starts
its coherent oscillations around the equilibrium value $\theta = 0$ of
a potential as in (3). However, because of the very small value (6) of the
effective
axion angle left after inflation, the contribution of axions produced by
the misalignment mechanism \cite{pww} to the present mass density of the
universe is
suppressed, due to the fact that it is proportional to the square of the
effective misalignment angle. As a result, it is no longer
necessary to lead
to the constraint $m_\alpha \grtsim 10^{-5}\rm eV$, ( $f_\alpha \lessim
10^{12}\rm GeV$ ).

For the present scenario, the value $f_\alpha \sim M_P$ is both possible and
consistent: the satisfied condition $H_* <
f_\alpha$ can prevent potential restoration of the Peccei-Quinn symmetry (by
the Hawking temperature) before the end of inflation \cite{lyth} and the also
satisfied condition $T_* < f_\alpha$ is necessary for not
restoring the symmetry after inflation. With $f_\alpha \sim M_P$ the axions
develop a very small mass $m_\alpha \sim 10^{-12}\rm eV$ and are precluded
from being the dark matter. Their couplings are also very
small. So, in this case the axions are invisible indeed.
\bigskip

\noindent {\bf Acknowledgements}
\medskip
\nobreak

It is a pleasure to thank D.Bailin, A.E.Everett and A.Liddle for discussions.


\newpage
\begin{thebibliography}{99}
\bibitem{pq}R.D.Peccei and H.Quinn, Phys.Rev.Lett. {\bf 38} (1977) 1440;
Phys.Rev. {\bf D16} (1977) 1791; S.Weinberg, Phys.Rev.Lett. {\bf 40}
(1978) 223;
F.Wilczek, Phys.Rev.Lett. {\bf 40} (1978) 279.
\bibitem{sikivie}P.Sikivie, Phys.Rev.Lett. {\bf 48} (1982) 1156.
\bibitem{vs}A.Vilenkin and E.P.S.Shellard, Cosmic Strings and other
Topological Defects (Cambridge U. Press, 1994).
\bibitem{invisible}J.E.Kim, Phys.Rev.Lett. {\bf 43} (1979) 103;
M.A.Shifman, A.I.Vainshtein and V.I.Zakharov, Nucl.Phys. {\bf
B166} (1980) 493; M.Dine, W.Fischler and M.Srednicki,
Phys.Lett. {\bf 104B} (1981) 199; A.P.Zhitnitskii,
Sov.J.Nucl.Phys. {\bf 31} (1980) 260.
\bibitem{kim}J.E.Kim, Phys.Repts. {\bf 150} (1987) 1; H.-Y.Cheng,
Phys.Repts.{\bf 158} (1988) 1; R.D.Peccei, in ``CP Violation'',
ed.C.Jarlskog (Word Scientific, 1989).
\bibitem{turner}M.S.Turner, Phys.Repts. {\bf 197} (1990) 67; G.G.Raffelt,
Phys.Repts. {\bf 198} (1990) 1; E.W.Kolb and M.S.Turner, The Early Universe,
(Adddison-Wesley, 1990).
\bibitem{pww}J.Preskill, M.Wise and F.Wilczek,
Phys.Lett. {\bf 120B} (1983) 127; L.Abbott and P.Sikivie,
Phys.Lett. {\bf 120B} (1983) 127; M.Dine and W.Fischler,
Phys.Lett. {\bf 120B} (1983) 137.
\bibitem{ve}A.Vilenkin and A.E.Everett, Phys.Rev.Lett. {\bf 48} (1982) 1867;
Nucl.Phys. {\bf B207} (1982) 43; T.W.Kibble, G.Lazarides and Q.Shafi,
Phys.Rev. {\bf D26} (1982) 435.
\bibitem{davis}R.L.Davis, Phys.Lett. {\bf B180} (1986) 225; R.L.Davis and
E.P.S.Shellard, Nucl.Phys. {\bf B324} (1989) 167; R.A.Battye and
E.P.S.Shellard, Phys.Rev.Lett. {\bf 73} (1994) 2954;
Nucl.Phys. {\bf B423} (1994) 260.
\bibitem{hs}D.Harari and P.Sikivie, Phys.Lett. {\bf 195B} (1987) 361;
C.Hagmann and P.Sikivie, Nucl.Phys. {\bf B363} (1991) 247.
\bibitem{rs}G.G.Raffelt, hep-ph/9502358; P.Sikivie, hep-ph/9503292.
\bibitem{dp}T.Damour and A.M.Polyakov, Nucl.Phys. {\bf B423} (1994) 532;
Gen.Rel.Grav. {\bf 26} (1994) 1171.
\bibitem{cev}C.E.Vayonakis, SUSX-TH/95-16 (April 1995).
\bibitem{guth}A.H.Guth, Phys.Rev. {\bf D23} (1981) 347; for a review, see
A.D.Linde, Particle Physics and Inflationary Cosmology (Harwood, 1990);
E.W.Kolb and M.S.Turner, The Early Universe (Addison-Wesley, 1990);
K.A.Olive, Phys.Repts. {\bf 190} (1990) 307.
\bibitem{ll}A.D.Linde and D.H.Lyth, Phys.Lett. {\bf B246} (1990) 353.
\bibitem{ls}G.Lazarides and Q.Shafi, Phys.Lett. {\bf B115} (1982) 21;
H.Georgi and M.Wise, Phys.Lett. {\bf B116} (1982) 123.
\bibitem{lps}G.Lazarides, C.Panagiotakopoulos and Q.Shafi,
Phys.Lett. {\bf B192} (1987) 323; S.Dimopoulos and L.J.Hall,
Phys.Rev.Lett. {\bf 60} (1988) 1899; G.Lazarides, R.Schaefer, D.Seckel and
Q.Shafi, Nucl.Phys. {\bf B346} (1990) 193.
\bibitem{pi}S.-Y.Pi, Phys.Rev.Lett. {\bf 52} (1984) 1725; M.S.Turner,
Phys.Rev. {\bf D33} (1986) 889; A.D.Linde, Phys.Lett. {\bf B201} (1988) 437.
\bibitem{tw}M.S.Turner and F.Wilczek, Phys.Rev.Lett. {\bf 66} (1991) 5.
\bibitem{twz}M.S.Turner, F.Wilczek and A.Zee, Phys.Lett. {\bf
B120} (1983) 127; M.Axenides, R.Brandenberger and M.S.Turner,
Phys.Lett. {\bf B128} (1983) 178; P.J.Steinhardt and M.S.Turner,
Phys.Lett. {\bf B129} (1983) 51; D.Seckel and M.S.Turner, Phys.Rev. {\bf
D32} (1985) 3178; L.A.Kofman and A.D.Linde, Nucl.Phys. {\bf B282} (1987) 55;
for a review, see A.D.Linde, ref. \cite{guth}; A.Liddle and D.H.Lyth,
Phys.Repts. {\bf 231} (1993) 1; D.Scott, J.Silk and M.White, Science
{\bf 268} (1995) 829.
\bibitem{linde}A.D.Linde, Phys.Lett. {\bf B259} (1991) 38.
\bibitem{dv}T.Damour and A.Vilenkin, IHES-P-95-26, hep-th/9503149.
\bibitem{cfkrr}G.D.Coughlan, W.Fischler, E.W.Kolb, S.Raby and G.G.Ross,
Phys.Lett. {\bf B131} (1983) 59; J.Ellis, D.V.Nanopoulos and M.Quiros,
Phys.Lett. {\bf B174} (1986) 176; O.Bertolami, Phys.Lett. {\bf B209}
(1988) 277; B.de Carlos, J.A.Casas, F.Quevedo and
E.Roulet, Phys.Lett. {\bf B318} (1993) 447; T.Banks, D.B.Kaplan and
A.E.Nelson, Phys.Rev. {\bf D49} (1994) 779; T.Banks, M.Berkooz and
P.J.Steinhardt, RU-94-92, hep-th/9501053.
\bibitem{gsw}M.B.Green, J.H.Schwartz and E.Witten, Superstring Theory
(Cambridge U. Press, 1987).
\bibitem{cd}M.Cvetic and R.L.Davis, Phys.Lett. {\bf B296} (1992) 316.
\bibitem{bbsms}T.Banks, M.Berkooz, S.H.Shenker, G.Moore and
P.J.Steinhardt, RU-94-93, hep-th/9503114.
\bibitem{gno}M.K.Gaillard, H.Murayama and K.A.Olive, UMN-TH-1334/95,
LBL-37019, UCB-95/109 (April 1995).
\bibitem{kls}L.Kofman, A.D.Linde and A.Starobinsky, Phys.Rev.Lett. {\bf
73} (1994) 3195; Y.Shtanov, J.Traschen and R.Brandenberger, Phys.Rev.
{\bf D51} (1995) 5438.
\bibitem{lyth}D.H.Lyth and E.D.Stewart, Phys.Rev. {\bf D46} (1992) 532.
\end{thebibliography}
\end{document}